\documentclass{aa}
\usepackage{graphicx}
\usepackage{natbib}
\def\m273{Mrk~273}

\def\a+a{A\&A} 
\def\subsun{\mbox{$_{\normalsize\odot}$}}
\def\kms{\,km$\,$s$^{-1}$}
\def\deg{\hbox{$^\circ$}}
\def\sun{\hbox{$\odot$}}
\def\nodata{\multicolumn{1}{c}{$\cdots$}}
\begin{document}
   \title{An embedded circumnuclear disk in Mrk~273}
   \author{Hans-Rainer Kl\"ockner
          \inst{1,2}
          \and
          Willem A. Baan\inst{2}
          }

   \offprints{Hans-Rainer Kl\"ockner}

   \institute{Kapteyn Institute, University of Groningen,
              P.O. Box 800, 9700 AV Groningen, The Netherlands \\
              \email{hrkloeck@astro.rug.nl}
         \and
             ASTRON, Westerbork Observatory,
             P.O. Box 2, 7990 AA Dwingeloo, The Netherlands\\
            }

   \date{Received 8 January 2004 / Accepted 13 February 2004}
   
%
%

\abstract{Radio observations using very long baseline interferometry
(VLBI) and the Westerbork interferometer have been carried out to
study the hydroxyl Megamaser emission in Mrk~273 at different spatial
resolutions. Line and continuum observations were carried out by the
European VLBI network (EVN) at 1.6 GHz and display a number of
distinct structural components in the central arcsec$^2$ region.
The observed continuum emission shows three prominent regions with
both flat and steep spectral indexes.\\
The hydroxyl (OH) emission detected by the EVN measurements accounts
for only 12~percent of the total OH emission in Mrk~273, but it does
show the same dominant 1667 MHz line emission components as the WSRT
observations.  The spatial distribution of the maser emission provides
a high resolution view of the molecular environment in the nuclear
region. The OH emission has only been detected toward a distinct radio
source in the northern nucleus with a spatial extent of 108~pc. The OH
emission is only partially superposed on the radio continuum and is
associated with the near-infrared emission source. The low pump
efficiency of the maser and the OH main-line ratio suggest that this
emission originates in an optically thin and unsaturated maser
environment with a complex pumping scheme that cannot be explained by
radiative infrared pumping with a single dust temperature. The
specific line emission pattern and the line-of-sight velocities
indicate the organized structure of an edge-on disk/TORUS with a
Keplerian rotation surrounding a central object with a binding mass of
$1.39\pm0.16\times10^{9}$~M\subsun .
}
\maketitle
%
%
\section{Introduction}
\label{sec:introduction6}

Extragalactic hydroxyl (OH main-lines) emission has been studied since
the early eighties \citep{1982ApJ...260L..49B}, establishing a new
class of extragalactic masers with unexpected isotropic luminosities
of several magnitudes higher than the most luminous galactic
counterparts (e.g. W3(OH)). The OH Megamaser (OH-MM) galaxies are a
sub-sample of the ultra-luminous infrared galaxies (ULIRG) that are
morphologically peculiar due to ongoing mergers. The nuclei exhibit
phenomena related to (circum-nuclear) starburst (SBN) activity and/or
the presence of an active galactic nucleus (AGN). The high molecular
and dust content of the nuclear regions together with the nuclear
activity render these sources ideal probes for studying the
circumnuclear environment. The exceptional line width of the OH
emission has been thought to exclusively trace the circumnuclear
environment close to a central power source \citep[e.g. see
][]{hrkjenam}. So far, only a small fraction of the prominent OH-MM
galaxies has been observed by using
very long baseline interferometry. These studies have shown a rather
complex picture of continuum and the maser emission inside the nuclear
region \citep[for a detailed list see][]{2002evn6.proc..175}.

The galaxy Mrk~273 (UGC~08696, IRAS~13428$+$5608) has been classified
as a ULIRG with an infrared luminosity of L$_{\rm FIR}$~=~$1.208\times
10^{12}$~L\subsun . Its warm infrared color (ratio of the 60~$\mu$m to
100~$\mu$m IRAS flux of 1.02) suggests that the bulk of the infrared
excess is associated with dust heated by nuclear star formation rather
than by a dominant contribution of an accretion disk \citep[][and
references therein]{1996ARAA..34..749S}. In the optical Mrk~273
displays a morphologically disturbed structure with a thin tidal tail
extending towards the south by 50~arcsec (1.3~arcsec $\sim$
1~kpc)\footnote{The optical redshift of z~=~0.03778,
q$_{0}$=~0.5~\kms\ and H$_{0}$=~75~\kms~Mpc$^{-1}$ are used for all
further estimates in this paper; therefore 1~mas corresponds to
0.7389~pc.} and a fan-like plume. The infrared excess, the distinct
morphology and the observed soft X-ray halo of
$\sim$50\arcsec~$\times$~30\arcsec\ indicate a merger event within the
last 10$^{8}$~yrs \citep{1997ApJ...490L..29K}. The fan-like plume
itself is 15\arcsec\ in size and has a rather complex structure
consistent with the inner interaction region in a galactic merger. The
central region of Mrk~273 harbors prominent dust lanes, ionized gas,
and several sources visible at different emission bands, which
complicates the diagnosis of the nuclear energetics. Strong optical
emission lines from two distinct regions indicate the presence of a
low-luminosity LINER nucleus in the north with a size of around
1~arcsec and a Seyfert~2 nucleus located 4~arcsec towards the
southwest \citep{1999ApJ...527L..13C}.

The LINER source in the north reveals both diffuse and compact radio
emission components, which support the notion of enhanced star
formation as well as the optical classification
\citep{2000ApJ...532L..95C}. On the other hand, hard X-ray emission
has only been detected towards this source, which indicates either a
heavily obscured high-luminosity AGN or a less obscured low-luminosity
AGN as a nuclear power plant \citep{2002ApJ...564..196X}. The
kinematical structure of this northern source has been well studied at
various resolutions using [O~III], HI, CO, and OH line emission. These
observations give evidence of a disk-like structure of less than 800
mas in size \citep{1999ApJ...527L..13C, 1988ApJ...329..142S,
1998ApJ...507..615D, 1987ApJ...321..225S}.

The southern Seyfert~2 source shows enhanced emission in the
near-infrared (NIR), only soft X-ray emission, and almost no emission
in the radio, which indicates either another dust enshrouded AGN or
the possible interaction region of an ionization cone of the northern
nucleus \citep{1997ApJ...490L..29K, 2002ApJ...564..196X}. Radio
observations of the central part of Mrk~273 reveal one additional
object to the south-east of the northern nucleus, which has no NIR
counterpart and no hard X-ray emission.  However, this source does
show a distinct jet-like morphology in the radio and displays faint HI
absorption and CO emission indicating a possibly outflow from one
additional nucleus into the surrounding ISM
\citep{1997ApJ...490L..29K,1998ApJ...507..615D, 2000ApJ...532L..95C}.

New WSRT and EVN observations of the OH line and the continuum
emission in the Megamaser galaxy Mrk~273 are presented in this
paper. The OH emission structure at the northern nucleus and the
continuum structure are used to understand the kinematics and the
physical properties of the environment in the nuclear region.

\section{Observations and data reduction}

The hydroxyl line- and continuum emission in Mrk~273 was observed
with the Westerbork Synthesis Radio Telescope [WSRT] and the European
VLBI Network [EVN]. The observations reveal structures spanning scale
sizes of about three orders of magnitude ranging from galactic to
nuclear scales.

The WSRT observations were made on 7~February~2002 with two hours of
on-source observations and respectively 3 and 4 min on the calibrator
3C~286 at the beginning and the end of the on-source observation. The
dual polarization line observations had a bandwidth of 20~MHz using a
rest frequency between the two hydroxyl main lines (1666~MHz) and a
heliocentric velocity of 11326~\kms. The set-up with 256 channels gave
a velocity coverage of about 3734~\kms\ and a velocity resolution of
about 14.6~\kms\ per channel. At the source distance, the OH emission
lines are red-shifted out of the protected band and into the band
where the GLONASS global positioning satellite system operates. As a
result considerable radio frequency interference (RFI) was encountered
in the measurements. The cross correlation products produced by the
WSRT interferometer were used to eliminate the influence of RFI on the
data.  The amplitude and gain calibrations were performed after
applying the Tsys measurements to the data. The RFI-affected data set
of the calibrator source was excised iteratively with an
automated calibration and flagging procedure based on individual tasks
from the Astronomical Imaging Processing System [AIPS].  After final
editing of the calibrator, the phases and the band-pass corrections
(frequency dependent gains) were used to calibrate the dataset of the
target source. Final RFI excising of the data of the target source was
performed using the same automatic excising method \citep[for a
detailed description see][]{2004hrkphd}. Observations of a few hours
with the east-west WSRT array result in a relatively poor UV-coverage
and detailed imaging was not carried out using this
dataset. After applying all corrections to the data set of the program
source, a Total-Power emission spectrum was produced from the
data in the UV plane (see Fig.~\ref{fig:wsrtspec}).

The EVN observations of Mrk~273 were made on 10 February 2000 using
seven antennae Effelsberg, Onsala [85~ft], Jodrell Bank [Lovell],
Medicina, Noto, Torun and the phased WSRT array. Due to weather and
technical problems only half of the scheduled 12~hrs observation have
been used. The data were processed with the correlator at the
Joint Institute for VLBI in Europe [JIVE] in Dwingeloo. To
observe both OH main-line emissions (1667 and 1665~MHz), the band pass
was centred at a mid frequency of 1666.38~MHz and assuming a
optical heliocentric velocity of 11350~\kms . The observations used a
total bandwidth of 8~MHz covering a velocity range of 1494~\kms\ in
two polarizations, 256 channels and 2~bit correlation, which leads to
a velocity resolution of 5.8~\kms\ per channel in the source rest
frame. The observation was performed in phase referencing mode,
using a 13~minute (10 + 3~minutes) cycle to cover the target source
Mrk~273 and the phase-calibrator source J~1337$+$5501. The projected
telescope spacings lead to a synthesized beam width of minimal
30~mas. Data reduction and analysis was performed using AIPS in
combination with routines from the Groningen Image Processing System
[GIPSY]. Before standard calibration of the UV-dataset were performed,
new accurate positions were applied for those telescopes that are not
part of the geodetic network \citep{2002evlb.conf....9C}. The
observational data was a-priori gain calibrated by using the system
temperatures measured at each individual telescope. The phases were
then calibrated by initial fringe finding followed by a full
self-calibration procedure on the phase calibrator. The bandpass of
the system was calibrated by using the phase-calibrator source,
because no simultaneous measurement at all telescopes of the scheduled
band pass calibrator (3C~286) had been recorded.  Further improvement
of the calibration could be achieved by performing an additional
self-calibration procedure on the channel with the highest line flux
and applying these phase corrections to the other channels. Imaging of
the line and continuum emission was produced after applying the
final calibration to all channels in the UV-dataset.

\section{Results}

\subsection{Radio continuum}

\begin{figure}[!h]
\centering \includegraphics[angle=0,width=7.5cm]{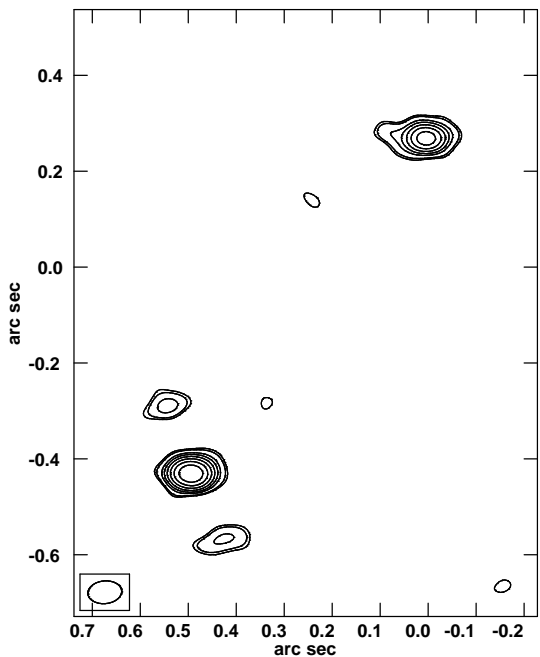}
\caption{Continuum emission structure of Mrk~273 at 1.6~GHz observed
on 10~February~2000 using the EVN network including the telescopes
Effelsberg, Onsala [85], Jodrell Bank [Lovell], Medicina, Noto, Torun
and the phased WSRT array. The measurements were mapped by
averaging the offline spectral channels, applying natural weighting,
and tapering, that leads to a spatial resolution of 71$\times $47 mas
oriented at 68.7\deg . The highest peak flux of 3.392~mJy per beam
corresponds to the southeast source.  The contour levels are in a
geometric progression of $\sqrt{2}$; hence every two contours
imply a factor of two in surface brightness. The first contour is 0.15
mJy beam$^{-1}$ corresponding to a 3.9$\sigma$ level.\vspace{0.8cm}}
\label{fig:273_cont} 
\end{figure}

%
The low-resolution (14\arcsec) continuum flux density at 1660 MHz was
obtained from the off-line emission in the WSRT spectrum to be
102~mJy.  This value is 13 to 22 percent lower than the catalogued
estimates of the NVSS and the FIRST databases at 1.4~GHz. Although the
catalogue images display unresolved point sources, they also show
differences in continuum flux densities on scale sizes between their
spatial resolution of 5.4\arcsec\ and 45\arcsec. This could indicate
the presence of a low brightness temperature continuum structure on a
scale size between these spatial resolutions. The flux discrepancy
with the WSRT observations could indicate a steep spectral index of
Mrk~273 or it is caused by the influence of RFI, which would lead to
an increase of the system temperature and a reduction of the continuum
level. The low resolution NVSS radio data was compared with the
catalogued infrared emission. The resulting q-value defined as log
(L$_{\rm FIR}$/L$_{\rm radio}$) of 2.28 is slightly lower than the
values typically seen in star-forming galaxies, which indicates that
any radio core and radio jet or lobe component only make a minor
contribution to the total radio emission of Mrk~273
\citep{2001ApJ...554..803Y}.

The complex continuum structure in the central region of 1~square
degree of Mrk~273 has been studied in previous observations.  The
continuum structure presented in Fig.~\ref{fig:273_cont} displays
multiple sources matching the known northern and south-eastern
structures in the nuclear region (natural weighted data set of
71$\times$47~mas resolution). The south-eastern component splits into
a marginally resolved triple source at position angle PA~$\sim$25\deg . In the
uniform weighted map at a slightly higher resolution of
41$\times$34~mas, the southern source of this triple structure reveals
two emission components, whereas the northern source remains
unresolved. The observed emission structures are consistent with
existing VLBA observations of the south-eastern region (1.3~GHz) at
somewhat lower resolution of 50~mas
\citep{2000ApJ...532L..95C}. Therefore, the EVN data partially trace
the organized structure of the southeast source comprising an
amorphous double jet of 370~mas in extent. The central source has a
brightness temperature on the order of 10$^{5}$~K, and is
characterized by a relatively flat spectral index ($\alpha $, F$\sim
\nu ^{\alpha }$) between 1.6 and 5~GHz of 0.26$\pm $0.05 \citep[5~GHz
data taken from][]{1997ApJ...490L..29K}. Similar values have been
found in a sample of ULIRGs, which suggests that synchrotron emission
could be the dominant emission process in such sources
\citep{1996ApJ...460..225C}.

\begin{figure}[!ht]
\centering
\includegraphics[angle=0,width=7.5cm]{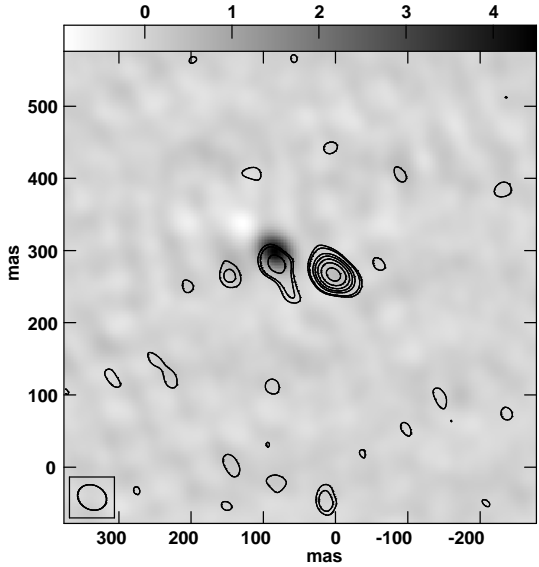}
\caption{Continuum emission superposed on the integrated OH line
emission observed in Mrk~273 by the EVN. A close-up of the northern
nucleus shows the continuum emission in contours superposed on the
integrated line emission in grey-scale color-coding. The dataset was
uniformly weighted, which results in a 41$\times$34~mas
resolution oriented at 64.3\deg . The contour levels are a geometric
progression in the square root of 2 starting at 0.15 mJy
beam$^{-1}$. Note that the line- and continuum emission displayed here
may not fully justify the scenario of the classical OH Megamaser
model, where the line emission would be located in front of the
continuum emission \citep{1989ApJ...338..804B}.}
\label{fig:273_contline}
\end{figure}

The continuum emission in the northern source of the nuclear region in
Mrk~273 shows an east-west elongated structure of 213~mas in size
(Fig.~\ref {fig:273_cont}). At the resolution of 41$\times$31~mas, the
northern radio emission (Fig.~\ref{fig:273_contline}) splits into two
components spatially separated by 77~mas in east-west direction. Each
component shows a brightness temperature on the order of 10$^{5}$~K.
The western component is somewhat brighter and slightly extended
towards PA~=~$-$33\deg , whereas the emission at the eastern component
fades towards the south. Similar characteristics for both continuum
components have been found with the VLBA~+~VLA at 1.3~GHz, which have a
slightly higher brightness temperature
\citep{2000ApJ...532L..95C}. Observations at 5~GHz with the MERLIN
array, with a resolution that is almost comparable to that of the
present EVN data, show enhanced continuum emission at the eastern
source as compared with the western source, which trend is also seen
in the NIR emission
\citep{2000ApJ...532L..95C,1997ApJ...490L..29K}. The spectral index of
both continuum sources is rather different between 1.6 and 5~GHz with
values of $\alpha _{\rm east}=1.5\pm0.4$ and $\alpha_{\rm
west}=0.01\pm0.12$, respectively. The flat spectrum suggests purely
thermal emission possibly caused by photo-ionization of the ISM by
star formation. While a steep spectrum at the eastern source would
indicate free-free absorption of the thermal emission or the
synchrotron emission, the radio continuum at these frequencies does
not provide clear clues of the nature of the nuclear engine
\citep{1992ARAA..30..575C}. Comparing the structures observed with the
EVN with images obtained by the VLBA~+~VLA array at 1.3~GHz shows that
these EVN observations, even with very poor UV-coverage at short
baselines, trace the inner part of the structures in the northern
nucleus of Mrk~273 \citep[see Fig.~2 in][]{2000ApJ...532L..95C}.  The
VLBA+VLA datashow both sources to be embedded in a diffuse emission
structure extending over 500$\times$300~mas.

\subsection{The line emission}

\begin{figure}[!ht]
\centering
\includegraphics[angle=0,width=7.5cm]{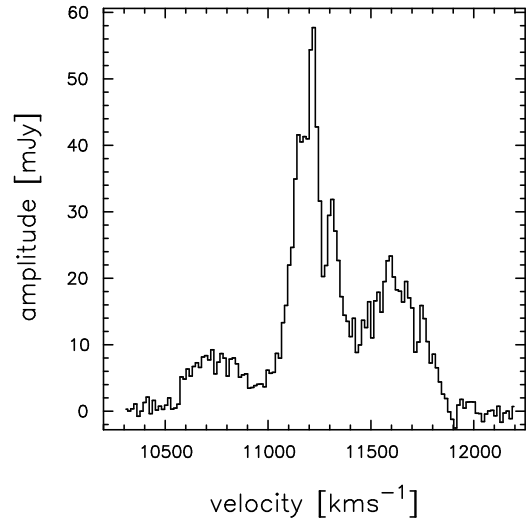}
\caption{Integrated line emission spectrum of Mrk~273 observed with
the WSRT. The spectrum has a velocity scale corresponding to a
heliocentric velocity of the 1667~MHz line and a spectral resolution
of 14.6~\kms . In order to compare this 20~MHz bandwidth observation
with the high resolution EVN spectrum of Fig.~\ref{fig:evnspecl}, the
velocity range has been cut off at 12250~\kms . A continuum flux
density of 102.6~mJy has been subtracted from the emission line
spectrum. The enhanced line feature at a central velocity of about
11326~\kms\ corresponds to the 1667~MHz main-line emission, whereas
predominately the emission at +365.6~\kms\ velocity offset is related
to the 1665~MHz main-line emission (see Fig.~\ref{fig:evnspecl}
and Fig.~\ref{fig:resispec}).
\vspace{0.0cm}}
   \label{fig:wsrtspec}
\end{figure}

\begin{figure}[!th]
\centering
\includegraphics[angle=0,width=7.5cm]{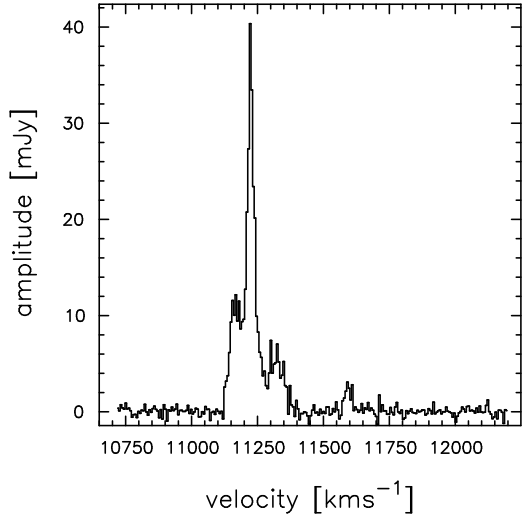}
\caption{Integrated line emission spectrum of Mrk~273 observed with
the EVN. The spectrum has a velocity scale corresponding to a
heliocentric velocity of the 1667~MHz line and a spectral resolution
of 5.6~\kms . Note that the velocity cutoff at higher velocities
corresponds to the cutoff in Fig.~\ref{fig:wsrtspec}. The strong line
features displayed here correspond to the 1667~MHz main-line emission
and the weak emission line is related to the 1665~MHz main line. The
expected velocity difference between the OH main-line transitions in
the velocity frame of the 1667~MHz line is +365.6~\kms\ at the redshift
of Mrk~273.}
\label{fig:evnspecl}
\end{figure}

It has become known that for extragalactic OH maser sources only a
small fraction of the total OH emission can be detected by
observations at parsec resolution. The high resolution EVN
measurements for Mrk~273 reveal only 12 percent of the hydroxyl
emission at low resolution obtained using the WSRT
(Fig.~\ref{fig:wsrtspec} and Fig.~\ref{fig:evnspecl}). A detailed
comparison of the individual line emission component presented in
Fig.~\ref{fig:resispec} shows that the central 1667~MHz line may
account for some 85\% of the EVN data, whereas the broader emission
features could not be recovered at all. The properties of the
individual line emission components have been listed in
Table~\ref{linepara}.

%
%
%
%

Extragalactic hydroxyl masers generally show dominant 1667~MHz
emission and are accompanied by weak (or sometimes non-existent)
1665~MHz line emission. At low resolution the OH spectrum shows three
features covering a total velocity range of 1257~\kms. Such
exceptional velocity ranges have also been seen in other Megamaser
galaxies that are characterized by a violent circum-nuclear
environment such as Arp~220, IRAS~14070$+$0525, and III~Zw~35
\citep[][]{1989ApJ...346..680B,
1992ApJ...396L..99B, 2001AA...377..413P}. The large velocity range of
the observed hydroxyl emission at WSRT scales complicates the
identification of individual line features with one of the hydroxyl
transitions. The theoretical velocity offset of both maser lines is
365.6~\kms\ at the assumed redshift of Mrk~273 of 0.03778 and in the
reference frame of the 1667~MHz line. Therefore, line features up to
about 11400~\kms\ will certainly be 1667~MHz features, whereas the
line features around 11600~\kms\ could be associated with the 1665~MHz
OH main line. The WSRT spectrum indicates a 1667/1665~MHz main-line
ratio of about 1.7, which is close to the 1.8
local-thermodynamic-equilibrium (LTE) value.

The line emission feature at the lowest velocity has been seen in the
early observation but has not been identified as such in the
literature\footnote{Not taking this particular line emission feature
into account in the observational set-up will cause systematic errors
for the kinematics and the OH content of low resolution investigations
of the hydroxyl emission in Mrk~273.} \citep[first detected
by][]{1992MNRAS.258..725S}. This component centred at 10773~\kms\ and
with a width of about 380~\kms\ complicates the interpretation of the
maser emission as it highlights gas at a non-systemic velocity. The
sub-structure of the main emission line at a velocity of around
11200~\kms\ shows three distinct components. The broad feature at
11600~\kms\ has a line width of 334~\kms\ (FWHM), which is
incidentally of the same order of magnitude as that of the broad line
at 10773~\kms .

The EVN measurements in Fig.~\ref{fig:evnspecl} show a less
complicated emission spectrum because some of the broad emission has
disappeared and the emission features can be easily associated with
the two hydroxyl transitions. The velocity range of the emission at
this resolution is 484~\kms, which is one third of the observed
velocity range at WSRT resolution. The spectrum displays strong
emission features at 11200~\kms\ associated with the 1667~MHz line and
a weak emission line at 11600~\kms. The clear triple structure of the
strong 1667~MHz emission changes dramatically as the fluxes of the f2
and f4 features are reduced at higher resolution
(Table~\ref{linepara}). The weak emission feature (f5) at 11600~\kms\
can be identified as 1665 MHz emission using the velocity offset of
369.9~\kms\ relative to the strongest 1667 MHz line feature f3.

\begin{table}
\caption{Properties of the individual emission line features obtained
by the WSRT and the EVN interferometers. The emission line features
are sorted by increasing centre velocity, corresponding to the
heliocentric velocity of the 1667~MHz emission line. The velocity
estimates have systematic errors of 14.6 and 5.8~\kms\, respectively,
that are related to the different spectral resolutions of the
measurements. A reduced chi-square fit has been used to estimate the
individual properties of the Gaussian shaped line profiles. The
theoretical velocity difference of the two OH main lines in the
reference frame of the 1667~MHz line is 365.6~\kms\ at the redshift of
Mrk~273.. The estimated velocity difference of the line features f3
and f5 match the theoretical offset within a systematic error of one
channel, indicating that they are a pair of hydroxyl main-lines.}
\label{linepara}{\scriptsize \rm 
\begin{tabular}[w]{lrrrr} \hline \hline
&  & \multicolumn{1}{c}{\bf WSRT} & \multicolumn{1}{c}{\bf EVN} &  \\ \hline 
&  &  &  &  \\
--~f1~component~-- &  &  &  &  \\ 
center velocity$_{\rm f1}$ & [\kms] & 10773.3 & \nodata &  \\ 
peak flux$_{\rm f1}$ & [mJy] & 7.84$\pm$0.57 & \nodata &  \\ 
FWHM$_{\rm f1}$ & [\kms] & 378.84 $\pm$36.05 & \nodata &  \\ 
L$_{\rm f1}$ & [L$_{\sun}$] & 106.99$\pm$12.78 & \nodata &  \\ 
&  &  &  &  \\ \hline
&  &  &  &  \\
--~f2~component~--&  &  &  &  \\ 
center velocity$_{\rm f2}$ & [\kms] & 11161.8 & 11168.0 &  \\ 
peak flux$_{\rm f2}$ & [mJy] & 39.34$\pm$1.00 & 11.63$\pm$0.43 &  \\ 
FWHM$_{\rm f2}$ & [\kms] & 128.64$\pm$4.44 & 51.79$\pm$2.75 &  \\ 
L$_{\rm f2}$ & [L$_{\sun}$] & 195.41$\pm$8.38 & 23.28$\pm$1.51 &  \\ 
&  &  &  &  \\ \hline
&  &  &  &  \\
--~f3~component~--&  &  &  &  \\ 
center velocity$_{\rm f3}$ & [\kms] & 11223.3 & 11224.4 &  \\ 
peak flux$_{\rm f3}$ & [mJy] & 33.74$\pm$1.74 & 33.93$\pm$0.67 &  \\ 
FWHM$_{\rm f3}$ & [\kms] & 44.4$\pm$2.70 & 29.9$\pm$0.77 &  \\ 
L$_{\rm f3}$ & [L$_{\sun}$] & 58.47$\pm$4.66 & 39.60$\pm$1.23 &  \\ 
&  &  &  &  \\ \hline
&  &  &  &  \\
--~f4~component~--&  &  &  &  \\ 
center velocity$_{\rm f4}$ & [\kms] & 11314.2 & 11299.8 &  \\ 
peak flux$_{\rm f4}$ & [mJy] & 25.80$\pm$1.16 & 5.24$\pm$0.28 &  \\ 
FWHM$_{\rm f4}$ & [\kms] & 90.97$\pm$5.57 & 114.86$\pm$11.23 &  \\ 
L$_{\rm f4}$ & [L$_{\sun}$] & 93.05$\pm$7.07 & 0.92$\pm$0.27 &  \\ 
&  &  &  &  \\ \hline
&  &  &  &  \\
--~f5~component~--&  &  &  &  \\ 
center velocity$_{\rm f5}$ & [\kms] & 11594.6 & 11594.3 &  \\ 
peak flux$_{\rm f5}$ & [mJy] & 19.76$\pm$0.60 & 2.71$\pm$0.51 &  \\ 
FWHM$_{\rm f5}$ & [\kms] & 334.56$\pm$14.16 & 30.88$\pm$6.75 &  \\ 
L$_{\rm f5}$ & [L$_{\sun}$] & 274.87$\pm$14.32 & 3.48$\pm$1.00 &  \\ 
&  &  &  &  \\ \hline
\end{tabular}
}
\end{table}

At a spatial resolution of a few milli-arcsec of the EVN data, the OH
emission has been found exclusively toward the northern source of the
nuclear region in Mrk~273 (see Fig.~\ref{fig:273_contline}). 

The OH emission structure has an extent of 146~mas and partially
covers the eastern continuum component. A marginal (3.5~$\sigma$) line
emission component at 11326~\kms\ and a line width of 54~\kms\ (FWHM)
has been detected toward the western continuum component. The centroid
of the line emission at the eastern source is displaced about 23.2~mas
towards the north relative to the continuum emission
(Fig.~\ref{fig:67vfield}). There is an asymmetry in the intensity of
the line features f2 and f4 that is most likely caused by the
association of feature f2 with the northern part of the continuum
emission serving as a background for maser amplification.  A
comparison of the properties of the individual OH\ line components in
table~\ref{linepara} shows that the EVN data accurately traces the
most compact and dominant OH emission seen towards Mrk~273.

\section{Discussion}

\subsection{Understanding the hydroxyl emission}

The hydroxyl emission in Mrk~273 displays characteristics that are
typical of extragalactic OH Megamaser sources, where at high spatial
resolution only a small fraction of the low resolution hydroxyl
emission is detected \citep{1998ApJ...493L..13L, 1999ApJ...511..178D,
hrk231nat}. The excitation mechanism of the OH molecules in Mrk~273 is
most likely a radiative pump provided by the infrared radiation field
similar to the conclusions for other OH-MM \citep{1989ApJ...338..804B,
1990AA...229..431H, 1997Natur.386..472S}.

Mrk~273 exhibits two distinct near-infrared emission regions at
arcsec-scales (south-west and north) with slightly stronger emission
towards the northern nuclear source where also the OH emission has
been detected \citep{1997ApJ...490L..29K}. By using the ratios of the
individual near-infrared measurements of the two sources in in
Mrk~273, the infrared excess seen towards the northern source is
estimated to be 5.96$\times 10^{11}$~L\subsun. Considering the
spectral width of the OH emission (e.g. r$_{\rm OH}$=~6.73~MHz) and
the applying this to the infrared emission (r$_{\rm FIR}$), a lower
limit of the OH pumping efficiency can be estimated using the
following ratio:
\begin{equation}
{\rm {P_{OH}}=\frac{{L_{OH}}\times {r_{FIR}}}{{L_{FIR}}\times {r_{OH}}}},
\label{effie}
\end{equation}

\noindent where L$_{\rm OH}$ and L$_{\rm FIR}$ are the
luminosities. The large scale OH emission in Mrk~273 shows a pump
efficiency of the order of 0.29 percent. An upper limit for the
efficiency of 0.40~\% can be found by using the maximum infrared flux
of the pump transition being closest to the black body peak and the
maximum peak flux of the observed OH line emission \citep[for the OH
transitions see: ][]{1977AA....60...55D}. Nevertheless, in both cases
the low pump efficiency indicates an unsaturated maser
process. Because the maser emission would be saturated if every
available pumping event (in this case a infrared photon) produces a
maser photon with an efficiency that depends only on the details of
the pumping scheme \citep[see page 81:][]{1992asma.book.....E}.
Saturation would occur when the measure of amplification by the maser
process directly affects the pumping efficiency. The ratio $\gamma$ of
the maser intensity at with saturation takes place in the 1667~MHz
hydroxyl line is about unity:
\begin{equation}
\centering
{\rm \gamma =\frac{I_{sat}}{I_{unsat}}\sim \frac{P_{OH}\Gamma}{A_{2^{-}2^{+}}}}
\label{satur}
\end{equation}

\noindent where I$_{{\rm sat}}$ is the intensity at saturation of the
maser, I$_{{\rm unsat}}$ is the unsaturated maser intensity,
A$_{2^{-}2^{+}}$ the Einstein coefficient of the 1667~MHz transition,
P$_{{\rm OH}}$ is the efficiency as defined above, and $\Gamma$ is the
loss rate. The loss rate $\Gamma $\ is generally of the same order as
the collision rate. For OH in HI regions, it has been shown that the
OH-ion interaction is the most important collision process, which
leads to a collision rate of $37\times 10^{-9}$~N$_{\mathrm{HI}}$,
where N$_{\mathrm{HI}}$ is the hydrogen density
\citep{1968ApJ...151..163R}. The hydrogen density is not generally
known but it can be estimated using the molecular hydrogen
abundance. The abundance of the molecular hydrogen has been
dynamically modeled for Mrk~273 to be 1860 H$_{2}$ molecules per
cm$^{-3}$ in the northern nucleus \citep{1998ApJ...507..615D}. The
relative abundance of HI to molecular hydrogen is assumed to be
similar to the galactic value of 20 \citep[valid for the inner 300~pc
in the Milky Way; ][]{1989IAUS..136...89G}. For a saturation ratio
$\gamma$~=~1, the natural logarithm of the gain threshold for maser
amplification in Mrk~273 as expressed in Eqs.~\ref{satur} is 10.8;
above this value saturated maser emission is expected. The saturation
parameter can be estimated observationally by determining the optical
depth form either the maser main-lines or one maser line in
combination with the continuum emission \citep[for an explanation see
][]{1990AA...229..431H, 1968ApJS...15..131G}. Since a clear separation
of the two OH main lines is not possible for the WSRT observation, an
upper limit on the gain of $-$0.43 can be estimated using the
continuum and the peak flux of $-$0.43 of the 1667~MHz line emission.
An additional constraint on the amplification process can be made
using the low resolution WSRT OH data and assuming that the radio
continuum observed in the northern source serves as a background for
the more diffuse OH emission, which is missed at the resolution of the
EVN. The continuum mission in the northern source in Mrk~273 has been
imaged in detail with the VLBA, showing a diffuse continuum structure
being punctuated by a number of compact sources (SNRs) with brightness
temperatures larger than 3~$\times 10^{6}$~K
\citep{2000ApJ...532L..95C}. The maximal brightness temperature of the
diffuse maser emission can be estimated from the missing line emission
features in the EVN data and the corresponding features in the MERLIN
observation \citep{2000MNRAS.317...28Y}. The diffuse maser emission
that is not seen by the EVN observations accounts for brightness
temperatures of the order of 4~$\times 10^{5}$~K. Within the classical
OH maser amplification model, these assumptions suggest a maximal gain
of about 2.01, which is significantly lower than the estimated
saturation threshold
\citep{1989ApJ...338..804B,1985Natur.315...26B,1982ApJ...260L..49B}.

A further constraint on the nature of the maser emission follows from
evaluating the level populations to estimate the main-line
ratios. Since the excitation mechanism of the OH molecules is most
likely radiative pumping, the integrated infrared emission field
determines the OH main-line ratio.  Assuming a spectral-temperature of
a single black-body or a single grey-body, the population in the OH
ground level ($^{2}\Pi _{3/2}$) follows from the population cascade of
the rotational and vibrational levels \citep{1977AA....60...55D}. By
solving the rate equations for the individual levels up to J~=~9/2, a
gray-body temperature of 63~K and a black-body temperature of 68~K
suggest a main-line ratio ranging from 1.98 or 3.00, respectively
\citep[note that for the gray body estimate additional infrared data
has been used from][]{2001AA...379..823K}. It has been mentioned that
the extreme kinematics in Mrk~273 will affect the diffuse OH main-line
emission and the main-line ratio, which may lead to systematic
errors. Nevertheless, the velocity difference of the individual f3 and
f5 features suggests that these are a line pair. At low resolution
their line ration is 1.71 (see table~\ref{linepara}), which is close
to the LTE value of 1.8 and can not be explained by an infrared
pumping scheme. At high resolution in the EVN data, the main line
ratio is 12.5 based on a clear identification of both OH lines, which
is a value that cannot be produced by the infrared radiation field
with a single spectral temperature. Therefore, in order to obtain a
detailed model of the observed line ratios, detailed modelling of the
radiation field and of the level populations needs to be employed,
which will be done elsewhere. The overall maser emission process can
thus be further constrained to be an unsaturated maser, because for a
saturated maser the process would only depend on the details of the
infrared pumping scheme.

The OH emission at EVN resolution as shown in
Fig.~\ref{fig:273_contline} only shows enhanced maser emission at the
eastern source, where the NIR emission is a factor 2 stronger relative
to the western source \citep{1997ApJ...490L..29K}. Since the dominant
pumping lines of the OH molecule fall in the NIR, a strong OH maser
versus FIR dependence could be expected. A similar connection has been
found for Arp~220, where a diffuse emission structure follows the NIR
intensity contours \citep{1984ApJ...279..541B}. The spatial
distribution of the continuum emission and of the OH emission suggests
that the diffuse continuum emission serves as a background for the
amplification. The difference in intensity of the f3 and f4 1667~MHz
line features is possibly caused by the difference in background
continuum at these locations. The location of the individual line
features with respect to this continuum emission follows from the
velocity field shown in Fig.~\ref{fig:67vfield}.

A clear discrimination of the OH main-lines and still small systematic
errors in determining the line ratios may lead to an accurate estimate
of the OH column density. Assuming that both main-line components f3
and f5 and the continuum originate along the same line of sight, the
optical depth of the OH can be estimated to be $\tau _{{\rm
unsaturated}}$~=~$-$3.20, which corresponds to a column density of
${\rm N_{OH}=4.58\times 10^{14}\,T_{ex}\,\,cm^{-2}}$ \citep[for
explanation see ][]{1968ApJS...15..131G}. An estimate of the optical
depth also follows from a comparison of the continuum emission and the
properties of f3 leading to the apparent optical depth of $-$4.30
\citep[the different ways of estimating the optical depth are
described in][]{1990AA...229..431H}. Combining these estimates for the
optical depth suggests a covering factor of 0.32, that accounts for
how much the OH emission clouds cover the underlying continuum
emission. Altogether, the pumping efficiency, the infrared emission,
the optical depth, and the estimate of the saturation threshold
indicate that the hydroxyl emission in Mrk~273 is spatially extended
between a thousand and a few tens of parsecs and is explained by an
unsaturated maser emission process.

\subsection{The nuclear kinematics}

\begin{figure} 
\centering
\includegraphics[angle=0,width=9.0cm]{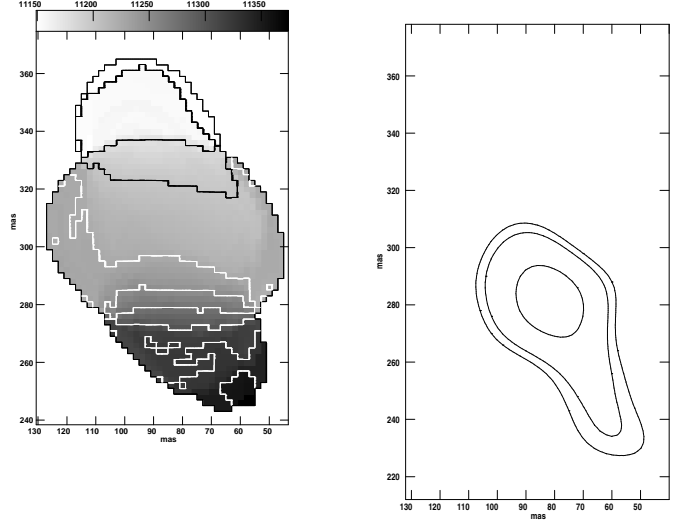}
\caption{The velocity field and the structure of the 1667~MHz OH
emission and the structure of continuum emission seen in a close-up of
the northern nucleus in Mrk~273. The spatial resolution is identical
to Fig.~\ref{fig:273_contline} and the velocity scale is the same as
in Fig.~\ref{fig:evnspecl}. The velocity field of the OH 1667~MHz
emission is shown in grey scale covering a velocity range between
11150 and 11350~\kms\ and contours are separated by 25~\kms . A
north-south velocity gradient of 1.62~\kms\ per mas is observed
(2.19~\kms\ pc$^{-1}$), where the northern side moves towards the
observer. Only the OH lines located in the centre and the southern
edge of the disk are superposed on the observed continuum emission.}
\label{fig:67vfield}
\end{figure}

The combination of OH data from the EVN and the WSRT traces the
hierarchical structure of the circum-nuclear environment from a few
tens of parsec up to kilo parsec scales in the northern nucleus of
Mrk~273. The kinematical pattern has been imaged using several
molecular tracers leading to a gas disk hypothesis for the northern
nucleus \citep{1999MNRAS.310.1033C,1998ApJ...507..615D}. In
particular, MERLIN imaging of the neutral hydrogen absorption at
spatial resolution of 200~mas shows a nuclear disk of about 800~mas in
extent. The kinematics of this disk is explained by solid-body
rotation with an velocity gradient of about 1.39$\pm $0.003~\kms\ per
mas (1.88~\kms\ pc$^{-1}$) in east-west direction
\citep{1999MNRAS.310.1033C}. On the other hand, MERLIN data of the OH
emission at a similar resolution displays a rather complicated picture
in which individual maser clumps do not provide any clear indication
of a disk structure at similar scale sizes
\citep{2000MNRAS.317...28Y}. At a resolution comparable to that of the
EVN observations (see Fig.~\ref{fig:67vfield}), the VLBA HI data show
that the northern nucleus is spatially resolved showing a 500$\times $
300~mas extended disk traced by HI in absorption with a slightly lower
velocity gradient of 1.5~\kms\ per mas (2.0~\kms\ pc$^{-1}$) east-west
direction than the OH and with an apparent flattening of the velocity at larger
radii \citep{2000ApJ...532L..95C}. The OH emission in the current EVN
data is unexpectedly different from the earlier data in that the
velocity gradient is north-south along the major axis of the emission
(Fig.~\ref{fig:273_contline}). Apparently the OH traces a
sub-structure in the nuclear region that has emission velocities
similar to those of the HI at lower resolution. At the location where
the OH emission indicates a clear north-south gradient (see
Fig.~\ref{fig:67vfield}), the HI absorption shows a flaring velocity
field associated with distinctly different kinematics.

\begin{figure}[h]
\centering
\includegraphics[angle=0,width=7.5cm]{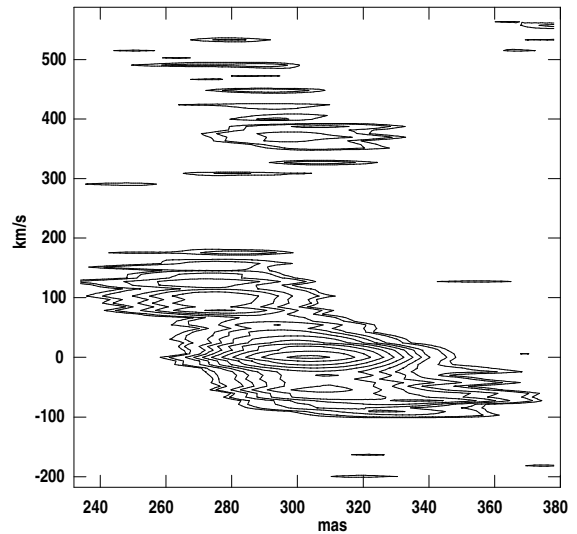}
\caption{The position-velocity structure of the hydroxyl emission in
the northern part of Mrk~273 observed with the EVN. The main-line
hydroxyl emission is displayed across the major axis of the eastern
sources at spatial resolution of 41$\times$34 mas (see
Fig.~\ref{fig:273_contline}).  The dominant emission structure in the
bottom part of the frame corresponds to the 1667~MHz line features.
The weaker emission in the upper part of the frame displays the
kinematics traced by the 1665~MHz emission. The velocity on the
vertical axis corresponds to the heliocentric velocity of the 1667~MHz
line emission. The declination refers to the central position, which
is consistent with all figures presented here. The centre velocity is
11222.3~\kms\ and the contours are in steps of 3 starting at 0.75 mJy
beam$^{-1}$.}
\label{fig:273_pv}
\end{figure}

The kinematical pattern of the OH line emission is shown in the
position-velocity (PV) diagram in Fig.~\ref{fig:273_pv} and displays
two almost similar velocity patterns for each of the OH main-lines
transitions. The features at the bottom of the PV-diagram represent
the three major components of the 1667~MHz line as seen in
Fig.~\ref{fig:evnspecl}. The velocity structure shows an organized
velocity field of Keplerian rotation around a central object with no
evidence of flattening at larger distances. The spatial extent and the
overall north-south velocity leads to a gradient of
1.62~$\pm$~0.37~\kms\ per mas (2.19~\kms\ pc$^{-1}$), which is of the
same order of magnitude as the MERLIN and VLBA estimates for the
east-west gradient in HI. The same velocity pattern repeats with
significantly lower intensity in the upper part of the PV-diagram,
where the 1665~MHz line is found.

The specific emission pattern of the OH emission (in particular of the
1667~MHz line) with a bright centre feature and two weaker components
at the edges is related to the maser amplification itself. We note
that the relative amplification of the maser emission is proportional
to the ratio of the OH column density over the OH velocity dispersion
\citep[e.g. see:][]{1995ApJ...440..619G}. In order to simulate the
representative emission pattern seen in the EVN spectrum in
Fig.~\ref{fig:evnspecl}, a 3-dimensional disk and torus geometrical
structure has been modelled in order to solve for the line-of-sight
maser amplification. In these geometries the amplification is
calculated on the basis of randomly distributed maser clouds within
the disk. To produces the smooth emission line spectrum that is common
for all OH Megamaser galaxies, each individual OH cloud needs to be
extended and exhibit internal velocity dispersion or turbulence. For a
107.9~pc (146 mas) extended structure and an enclosed mass on the
order of $1.39\pm 0.16\times 10^{9}$~M\subsun , the kinematics of a
solid body rotation, suggested by the HI absorption measurements
\citep{1999MNRAS.310.1033C}, does not account for the spectral
signature and the velocity range traced by the observed line features
f2, f3, and f4. The specific triple structure of the OH 1667~MHz line
emission can only be reproduced by either a disk or a torus structure
with Keplerian kinematics and seen almost edge-on with deviations in
the inclination of less then 10\deg\ \citep[for the dependence of the
spectral signature of maser emission on geometry and the
line-of-sight, see][]{2004hrkphd}.

Therefore, the OH emission line traced by the EVN observations is
explained by an almost edge-on disk or torus with Keplerian kinematics
placed at PA=~$-$15\deg\ of the eastern source in the northern nucleus
of Mrk~273.

At larger scale sizes the MERLIN data of the OH emission show no clear
kinematical pattern.  On the other hand, the WSRT spectrum
(Fig.~\ref{fig:wsrtspec}) shows an emission feature (f1) blue-shifted
by about 450~\kms . The presence of this feature could indicate a
more disturbed environment and may explain the complicated emission
pattern in the MERLIN data.  Because of the large velocity range of
the OH emission this feature did not get recognized in earlier
observations. Since no imaging data of this OH feature is available,
only an indirect comparison with optical spectroscopy could be made.
The OIII line emission shows two distinct lines with different
energetics at the northern nucleus \citep[note that the spatial
resolution of the reported observations is about 0.9~\arcsec
,][]{1999ApJ...527L..13C}. One OIII component indicates a highly
ionized gas phase with a rather disturbed velocity field and velocity
components predominantly 600~\kms\ blue-shifted. This velocity shift
is almost in agreement with the velocity difference seen for the OH
feature f1 traced by the OH emission. The other OIII line component
indicates low-excitation gas that traces a rather organized velocity
field oriented in east-west direction. The orientation and the
velocity range of about 2400~\kms\ indicate that this line traces
nuclear kinematics similar to those derived by radio observations at
comparable resolution \citep{1999ApJ...527L..13C,1999MNRAS.310.1033C}.

%
%
\begin{figure}[!ht]
\centering
\includegraphics[angle=0,width=7.5cm]{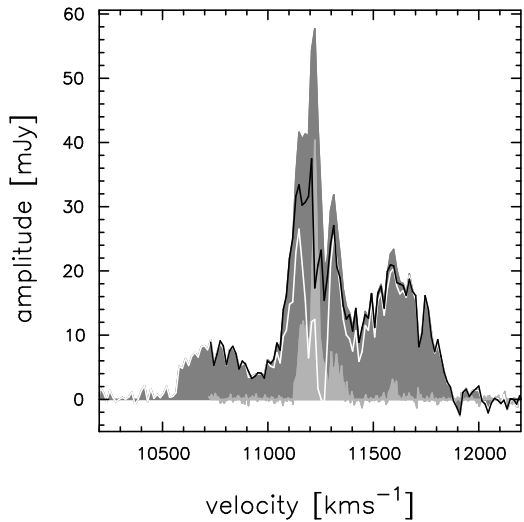}
\caption{A comparison of the integrated line emission spectra of the
hydroxyl emission observed in Mrk~273. The dark gray shaded region
displays the WSRT spectrum and the light gray shaded region the EVN
spectrum. The solid black line displays the difference between the
WSRT and the EVN measurements, while neglecting the different spectral
resolution of these observations. The solid white line represents the
difference of both observations, where the EVN spectrum, with a
spectral resolution of 5.6~\kms , is smoothed to the WSRT resolution
of 14.6~\kms .\vspace{0.0cm}}
\label{fig:resispec}
\end{figure}
%
%

The OH emission seen in Mrk~273 is possibly associated with both the
high and the low energy gas phase in the northern nucleus. The broad
blue-shifted OH and optical emission lines are most likely associated
with a starburst driven molecular outflow. Such a scenario has been
proposed and observed in other OH Megamaser sources such as Arp~220 or
III~Zw~35 \citep{1989ApJ...346..680B}. If a molecular outflow takes
place, the OH molecules must be contained in some dusty layer
expanding from the nuclear starburst traced in the radio continuum
\citep{2000ApJ...532L..95C}. The symmetry of the residual spectrum
seen in Fig.~\ref{fig:resispec}, where the EVN emission is subtracted
from the WSRT emission spectrum would then account for the diffuse OH
component on scale sizes of a few hundreds of parsec. The kinematics,
the source extent, and the enclosed mass estimates of the HI
absorption lines can be used to model the line-of-sight amplification
of the maser clouds embedded in a circular layer and their spectral
signature. Such a setup results into two distinct spectral line
features at both edges of the spectrum, each having a velocity
dispersion of a few hundreds \kms\ and covering a similar velocity
range as the residual OH emission
(Fig.~\ref{fig:resispec}). Therefore, the OH spectrum at velocities
higher than systemic would show a superposition of both OH main-lines
(see Fig.~\ref{fig:wsrtspec}) and would explain the complicated OH
velocity pattern seen at MERLIN resolution.\\

\section{Conclusions}

The flux density of the central line emission features of the
single-dish emission of Mrk~273 has been mostly recovered in a region
of around 108~pc at the eastern of the northern radio components. The
distinct velocity pattern and spectral shape of the maser spectrum
indicates that the emission originate in an almost edge-on disk or
torus covering the radio source. Observational data from various
wavelength regimes do not yet provide a completely consistent picture
of the nature of the nuclear power plant in the northern nucleus.

In the infrared, the total extent of the northern nucleus can be
determined by either a single gray-body emission area of about 418~pc
or a blackbody emission area of about 340~pc. These estimates are
remarkably consistent with the maximum extent of 370~pc found for the
radio continuum by the VLBA array and suggest that star-formation is
taking place on these scale sizes \citep{2000ApJ...532L..95C}. In
addition, EVN observation and other at higher resolution show a
distinct region of enhanced continuum emission located in the central
part in the northern nucleus of maximal 157~pc in extent. The combined
presence of these two distinct morphologies indicates that both a
starburst and an active galactic nucleus are present in the northern
nucleus of Mrk~273, which is optically classified as a LINER~1. In
addition, the northern nucleus shows exclusively hard X-ray emission
indicating either a heavily obscured high-luminosity AGN or a
low-luminosity AGN that mimics a LINER~1 spectrum with
photo-ionization due to hard photons \citep{2000ApJ...539..161T}. The
HI column density that is required to block the entire soft X-ray
emission towards the northern region is 4.1$\times$~10$^{23}$~cm$^{-2}
$, which is quite similar to estimates in the mid-infrared bands of
$\sim 5\times $~10$^{23}$~cm$^{-2}$
\citep{2002ApJ...564..196X}. Nevertheless, the HI column density seen
towards the individual continuum components in the northern region is
significantly lower with a value of
T$_{s}\times(1.8\pm0.3)$~10$^{20}$~cm$^{-2}$. These estimates could
only be similar for an unlikely spin temperature of about 2300~K,
which is an order of magnitude higher that the estimates of spin
temperatures based on a galactic gas-to-dust ratio and an optical
extinction similar to those detected in NGC~4945 or SgrA$^{\ast}$
\citep{2002ApJ...564..196X,2000ApJ...532L..95C}. On the other hand,
the OH column density N${\rm _{OH}=4.58\times
10^{14}\,T_{ex}\,\,cm^{-2}}$ observed towards the north-eastern source
and the standard galactic HI to OH abundance ratio implies a low
line-of-sight obscuration \citep{1980Natur.283..721M}. Therefore, the
OH maser emission combined with the steep radio spectral index
indicates the presence of a low luminosity AGN seen through a highly
ionized screen at the northern nucleus. The binding mass of the
central object in the north-eastern component is determined by
Keplerian kinematics and is on the order of $(1.39\pm 0.16)\times
10^{9}$~M\subsun . This is two orders of magnitudes larger than the
binding mass traced by the H$_2$O Megamaser emission in the galaxy
NGC~4258 \citep{1995ApJ...440..619G}.

\begin{figure}[tbp]
\centering
\includegraphics[angle=0, width=7.5cm]{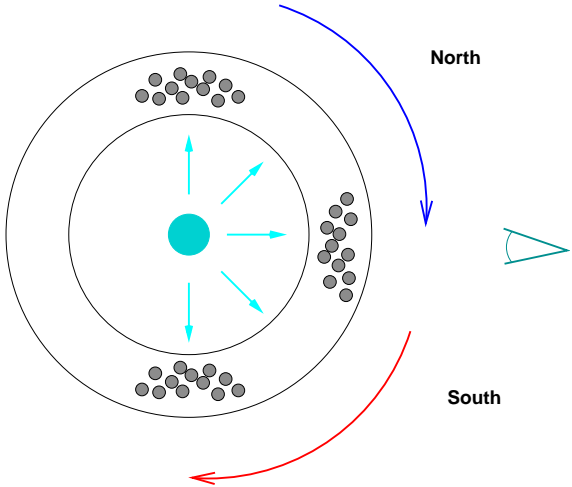} 
\caption{Schematic view of the maser emission seen with the EVN
observation. The relative amplification of the model maser emission
depends on the ratio of the OH column densities to the OH velocity
dispersion. At the top and bottom of the nuclear disk, the
line-of-sight column densities are relatively high with respect to the
centre, which is compensated by specific rotation velocities and
different line-of-sight dispersions. This scenario accounts for the
weaker and broader line features f$_2$ and f$_4$ and the central
strong narrow line feature f$_3$ of the 1667~MHz emission in
Figure~\ref{fig:evnspecl}.}
\label{fig:model}
\end{figure}

The spatial structure, spectral signature, and velocity pattern of the
OH emission in Mrk~273 at resolution of a few tens of parsec clearly
reveal an edge-on disk with Keplerian kinematics in the northern
nucleus. The distinct velocity pattern of the disk would represent a
kinetically independent structure in this nucleus, which does not
agree with the presence of an east-west elongated nuclear disk or
torus as traced by HI in absorption in the northern region
\citep{1999MNRAS.310.1033C}. However, the observed OH velocity
gradient is comparable to the gradient of the HI absorption, which
shows evidence for solid-body rotation and appears non-consistent with
the other evidence. Alternatively, this component could also be
explained by radial outflow.

The missing maser emission between the WSRT and the EVN observations
accounts for 80 percent of the total OH emission in
Mrk~273. Corresponding line features in the MERLIN observations show
that the OH emission traces a complex and disturbed velocity structure
at larger scale sizes \citep{1997ApJ...490L..29K}. The question
remains whether the velocity field of the large scale HI absorption
signifies rotation or radial outflow.  Alternatively the western
component of the northern nucleus, with its marginal narrow OH
emission component, its broad HI absorption feature, and its thermal
spectral index, may be interpreted as a region with enhanced star
formation rather than a second nucleus \citep{2000ApJ...532L..95C}.

The favored scenario for the northern nucleus is a combination of
star-formation that produces starburst driven winds, resulting in high
offset line features, and a low luminosity AGN obscured by a dusty
disk of 108~pc in extent producing a highly disturbed gas region that
mimics a LINER spectrum.  It should be noted that the hydroxyl maser
emission in Mrk~273 shows no saturation effects at kilo parsec and
parsec scale sizes. This is in contrast with the compact maser
components discovered in high-resolution observations of other bright
OH Megamaser sources, such as Arp~220, IRAS~17208$-$0014, and
III~Zw~35 \citep{1998ApJ...493L..13L,1999ApJ...511..178D}. In there
sources saturation and non-saturation has been concluded on the basis
of the compactness of the OH emission with respect to the
continuum. The current EVN observations show for the first time that
compact OH emission does remain unsaturated. It has also been shown
that the total OH maser emission at galactic size scales results from
an unsaturated maser process working with a pumping efficiency of only
a few one-thousands. Such low conversion efficiencies have been found
for most of the OH Megamaser galaxies, which indicates that saturation
effects may not play a significant role after all and support a
Megamaser model based on amplification of background radio continuum
by foreground molecular gas
\citep[][]{1985Natur.315...26B,1989ApJ...338..804B, 2004hrkphd}.\\

%
%

\begin{acknowledgements}

The European VLBI Network is a joint facility of European, Chinese,
South African and other radio astronomy institutes funded by their
national research councils. The Westerbork Synthesis Radio Telescope
is operated by the ASTRON (Netherlands Foundation for Research in
Astronomy) with support from the Netherlands Foundation for Scientific
Research (NWO). The authors thank M. Elitzur for fruitful discussions
on maser emission.

\end{acknowledgements}

%
%

%
%

\end{document}